# Real and imaginary part symmetries (RIPS) and artifacts removal in $^1$H magnetic resonance spectroscopy without water suppression

Zhengchao Dong[1,2]

1. Department of Psychiatry, Columbia University, College of Physicians and Surgeons, New York, NY, USA

2. New York State Psychiatric Institute, New York, NY, USA.

Correspondence to:

Zhengchao Dong, Ph.D.

Department of Psychiatry

Columbia University

New York, USA

Email: zhengchao.dong@nyspi.columbia.edu

Tel. (646) 774-5828

## ABSTRACT

### Purpose:

Proton MR spectroscopic imaging ($^1$H MRSI) without water suppression (WS) possess some distinct advantages over the conventionally used $^1$H MRSI with WS. However, the sideband artifacts in the non-water suppressed spectra hinder the applications of the $^1$H MRSI without WS. Although many hardware or software techniques to tackle the sidebands have been developed, they suffer from various shortcomings, such as the prolonged data acquisition time, insufficient elimination of the sidebands, or loss of some spectral information. Here, we present a software method that allows the elimination of sideband artifacts from $^1$H MRSI data acquired without WS.

### Methods

The presented method is based on the following observations: (1) the real part of the sidebands is anti-symmetric, and the imaginary part of sidebands is symmetric about the unsuppressed water signal, which is at the center of the spectrum, and (2) there is no observable metabolite resonances in the downfield region spectrum acquired with conventional MRSI pulse sequences. We model the complex sidebands in the downfield region using a singular value decomposition-based method, reconstruct from them the upfield sidebands according to the symmetry features of the sidebands, and subtract the sidebands from the spectrum. We demonstrate the method using phantom and in vivo data.

### Results

The method can remarkably reduce sideband artifacts in both the real and the imaginary parts of the spectrum. In both phantom data and high-quality in vivo data, the residual sidebands are smaller than those in the WS spectra.

### Conclusion

The method allows to remove sideband artifacts in both the real and the imaginary parts of the spectrum. The performance of the method on phantom data and high-quality in vivo data outperforms on poor quality in vivo data.

Keywords:

$^1$H MRSI, without water suppression, sideband artifacts, phase symmetry

## 1. INTRODUCTION

The water signal plays an important role in applications of proton MR spectroscopy ($^{1}$H MRS). The water signal is often used as an internal reference for absolute quantification of metabolite concentrations because of its relative stability in many physio-pathological conditions, its high signal to noise ratio (SNR), and its known or easily measurable concentrations.[1-3] It is also used for corrections of spectral imperfections, such as lineshape distortion, frequency shift, etc. to facilitate more accurate and/or precise spectral fitting.[2] In $^{1}$H MRS-based thermometry,[4, 5] the frequency of water signal, which is linearly dependent on the temperature in the physiological range, is measured with respect to temperature-independent frequencies of metabolites and used for the measurement of absolute temperature[6, 7]. Simultaneous measurements of metabolite concentrations and absolute temperatures of the brain using single voxel $^{1}$H MRS have been used in basic research fields and clinical settings.[8, 9]

Despite its usefulness, however, the water signal is all most exclusively suppressed in routine $^{1}$H MRS scans. This is because the $^{1}$H MR spectrum acquired without water suppression (WS) has sideband artifacts produced by the varying magnetic fields induced by mechanical vibrations of the exciting gradient coils.[2] These sidebands, with magnitudes comparable to and frequencies overlapping with those of the metabolite signals, hamper the spectral fitting and metabolite measurements. Therefore, the water signal is either not acquired at all or acquired in a separate scan before or after the water-suppressed $^{1}$H MRS scan for metabolite signal. Although a separate scan without WS only slightly increases scan time for single voxel $^{1}$H MRS, it will double the scan time for $^{1}$H MR spectroscopic imaging (MRSI), and therefore, it is impractical for clinical applications. Additionally, when the water signal is separately acquired, its effectiveness and usefulness are often compromised. For example, the mismatch of lineshapes between water suppressed and water unsuppressed signals due to subject motion will result in suboptimal corrections of effects of eddy current and inhomogeneous filed; the relative frequency shifts between separately measured signals of water and metabolites due to field shift will incur systematic errors in temperature measurement in $^{1}$H MRS-based thermometry.[7] On the other hand, the usefulness of water signal will be lost if it is not acquired.

The characteristics and mathematical descriptions of the sideband artifacts in the NWS $^{1}$H MRS are crucial for the development of methods for the NWS $^{1}$H MRS and have been documented in numerous publications.[2, 10-13] The two best-known, major features of the sidebands are that (1) their phases are associated with both the phases of the water signals and the polarities of the gradients, and (2) the sidebands are antisymmetric with respect of the water resonance in the frequency domain. These characteristics of the sideband artifacts have directly led to the development of data-acquisition-based techniques for the NWS $^{1}$H MRS.[12, 13] On the other hand, the generation of the sideband artifacts by the effects of the time-varying gradient fields is described as a multiplication of a phase term with the sideband-free $^{1}$H MRS signal. From this description, the non-water-suppressed (NWS) $^{1}$H MRS signal is further described as the summation of the sideband-free signal and the sideband signals[2]. These mathematical descriptions have been the basis for the development of postprocessing-based methods for the removal of sidebands in NWS $^{1}$H MRS.[2] We briefly review the two categories of methods in the next paragraphs.

The data acquisition-based techniques[12-15] were inspired by the correlations of the phases of sideband artifacts with the phases of the water signal or the polarity of the gradients. Examples of these techniques include (1) gradient cycling, which acquires pairs of NWS data whose sideband artifacts are in opposite phases due to opposite gradients, (2) water cycling, which acquires pairs of NWS data of opposite sidebands with opposite water signals, while keeping metabolite signal intact, and (3) metabolite cycling, which is the counterpart of the water cycling. While these experiment-based methods are very robust in sideband removal, a common problem is that they require modification of the pulse sequences and their parameters and are, therefore, not readily available on commercial MR scanners for a wide range of users. More problematically, these data acquisition-based techniques entail two interleaved acquisitions, e.g., one with and one without inversion of gradients, metabolite, or water signals. This is often unaffordable in clinical settings. Data undersampling techniques, such as SENSE[16] and compressed sensing,[17] that reduce the amount of k-space data points are used to compensate the increase of the scan time of these two-scan schemes. Therefore, these schemes avoided doubling the scan time at the expense of halving the SNR and did not change the double-scan nature of these NWS MRS techniques.

Some reference and postprocessing-based methods[18-20] acquire reference MRS data from a water phantom with identical experimental parameters as those of in vivo scans, which include pulse sequence, sequence timings, and locations of region of interest with respect to the iso-center of the gradients, etc. The information of the sideband artifacts in the reference signals is extracted and used to remove sideband artifacts of the NWS in vivo signals. The reference signals can be acquired during off-hours, mitigating the impact of additional scan-time. Therefore, the scan time penalty suffered by the data acquisition-based methods can be largely avoided. However, the performance of the reference-based methods may be limited by the mismatching of the linewidths and the lineshapes between the in vivo and the reference signals.[2, 21]

Two reference-free, postprocessing-based methods were proposed. One of the methods[22] used the property of the sidebands that the first order sidebands are anti-symmetric about the water signal and the fact that the downfield region is metabolite signal-free because the pulse sequence only selectively excite metabolite signals in the upfield. The method first extracts downfield sideband signals, inverts their phases, and reflects their frequency to the upfield; and it then subtracts the reflected and inverted sidebands from the original signal to remove the sidebands in the upfield region. The advantage of this method over the experimental and the reference-based methods is that it does not need sequence modification or a reference scan. But the method was only applied to phantom data, and the residues of the sidebands were clearly visible for signals at short TEs. More importantly, the theoretical basis of the method was incomplete: the anti-symmetric property of the sidebands is *only* true for the real part of the signal. In fact, the imaginary part of the sidebands is *symmetric* about the water signal, which is opposite to the real part of the sidebands. Therefore, the method will only remove the sidebands in real part but will *double* the sidebands in imaginary part spectrum. This will compromise either the accuracy or the precision of the metabolite measurement, depending on whether the full, complex data or only the real part is used in the spectral fitting (Supplementary Materials).

The other reference-free method is the modulus method,[11] which takes the modulus of the complex, NWS signal to reduce the sidebands. This approach is equivalent to abandoning the imaginary part of the signal and using only the real part of the signal. As a result, the random error of spectral fitting will increase by √2 if the fitting is performed on the full-time domain or frequency domain signal; If only the upfield or downfield signal in the frequency domain is used in the fitting, the random error will increase by a factor of 2.

The purpose of the present work is to propose a postprocessing-based method for removing sideband artifacts in NWS $^1$H MRSI. The method is hardware/sequence-independent and reference-free. The method exploits the symmetries of the real and imaginary parts of the sideband artifacts for the artifacts removal based on the following experimental and theoretical observations: (1) the real part of the sidebands is *anti-symmetric*, but its imaginary part is *symmetric* with respect to the water resonance, (2) the metabolite signals in the downfield region are invisible in sequences with limited and shifted excitation band, and (3) the sidebands in the upfield and downfield regions are related by a negative conjugate transform. We bud the method as real and imaginary part symmetry-based artifact removal (RIPSAR or RIPS for short) for $^1$H MRS without water suppression. In the following, we will first provide a background, a theoretical framework, and implementation procedures for the method, and then we will demonstrate the performance of the proposed method using phantom and in vivo data.

## 2. THEORY and METHODS

### 2.1 THEORY

We let $S_{w,0}(t)$ and $S_{m,0}(t)$ be the sideband-free, complex water and metabolite signals, respectively, and the non-water-suppressed proton MRS signal, $S_{nws}$, is described as:[2]

$$\begin{aligned} S_{nws}(t) &= [S_{w,0}(t) + \textstyle\sum_m S_{m,0}(t)]\, e^{i\vartheta_G(t)} \\ &= S_{w,0}(t) + S_{w,sa}(t) + \textstyle\sum_m S_{m,0}(t) + \sum_m S_{m,sa}(t) \end{aligned} \qquad (1)$$

where $S_{w,sa}$ and $S_{m,sa}$ are the sideband artifacts caused by the effects of the phase factor $e^{i\vartheta_G(t)}$ of the exciting gradients on the original *water* signal and the metabolite signals, respectively. However, the $S_{m,sa}$'s are more than 3 orders of magnitude smaller than the original metabolite signals and are not further considered. The sideband signal of the water is decomposed into the downfield (*df*) and the upfield (*uf*) parts:

$$S_{w,sa}(t) = S_{w,sa}^{df}(t) + S_{w,sa}^{uf}(t) \qquad (2)$$

Previous publications documented that the sidebands were anti-symmetric about water resonance.[10, 11] However, this description is only valid for the real part of the sidebands. In fact, the imaginary parts of the sidebands are symmetric (Fig. 1). Some metabolites, such as NAA (7.82 ppm) and Cr (6.65 ppm), have resonances in the downfield region;[23] but those resonances are deliberately not excited in routine $^1$H MRS studies using the limited bandwidth and appropriately adjusted center frequencies of the RF pulses, and they are not visible (Fig. 2). Therefore, the downfield region is virtually *signal-free* in water suppressed spectrum, and the peaks in the downfield region are solely the sideband artifacts, $S_{w,sa}^{df}(t)$ (Figs. 1 & 2).

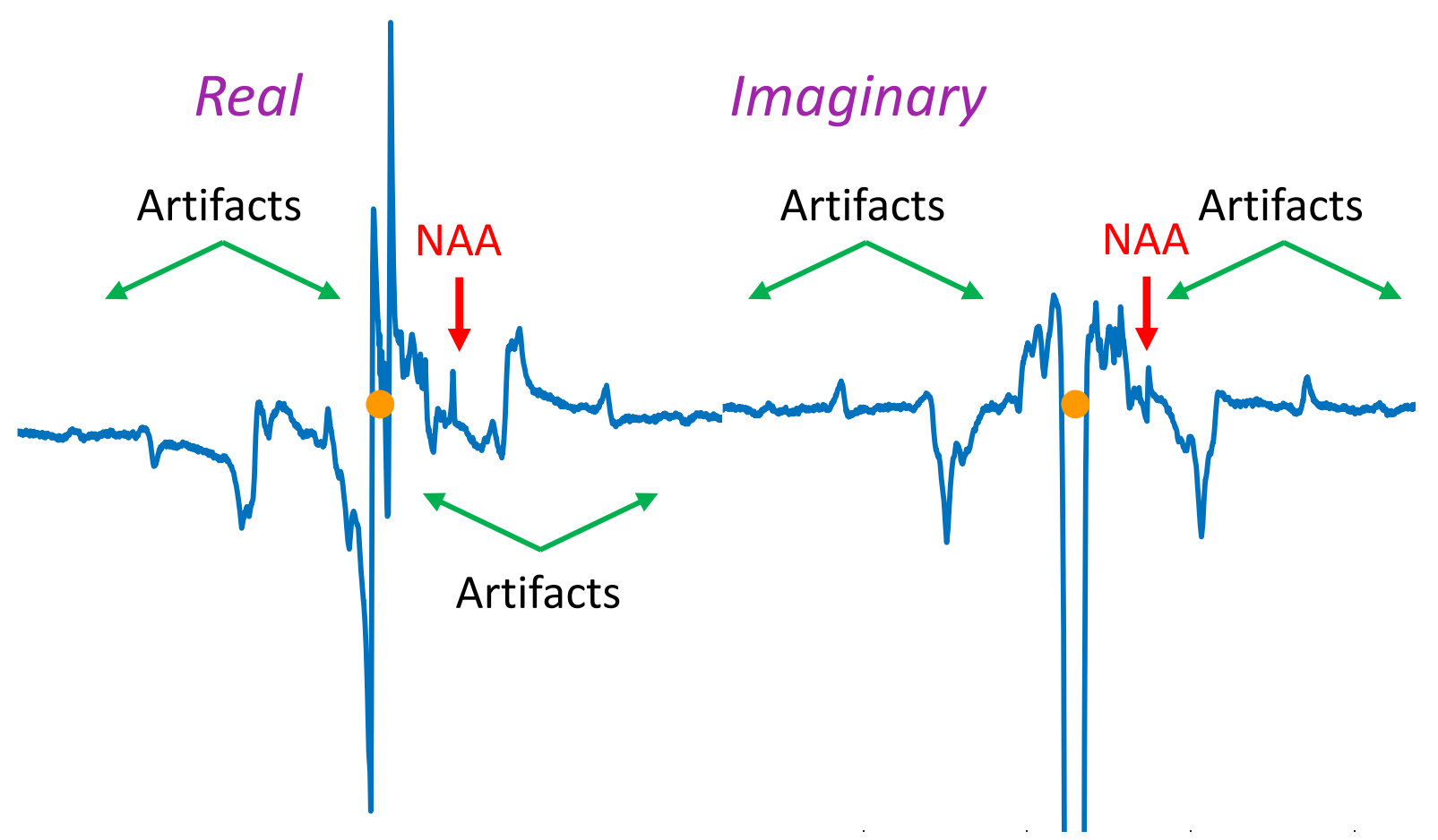

Figure 1. Real and imaginary parts of the sideband artifacts. The spectrum was obtained from an MRS phantom (The "Braino", GE Healthcare) without water suppression. The water signal was removed in the postprocessing for the exhibition of the sidebands. The real parts of the sidebands in the downfield and upfield regions about water resonance are anti-symmetric, and the imaginary parts of the sidebands are symmetric about the water resonance, whose position is shown by the orange dot.

However, the sideband artifacts in the upfield, $S_{w,sa}^{uf}(t)$, overlay with metabolite signals $\sum_m S_m(t)$, and they cannot be isolated and exacted from spectral fitting; but $S_{w,sa}^{uf}$ can be obtained from the negative conjugate of the $S_{w,sa}^{df}$:

$$S_{w,sa}^{uf} = -S_{w,sa}^{df}{}^{*} \quad (3)$$

where the "*" sign stands for complex conjugate transform that symmetrically reflects the frequencies of the sidebands from downfield (with positive frequency) to upfield (with negative frequency), and the negative sign "-" negates the phases of the real part of the sidebands and retains the phases of the imaginary part of the sidebands in the downfield region. Now the artifact signal in the entire signal is given as:

$$S_{w,sa}(t) = S_{w,sa}^{df}(t) + S_{w,sa}^{uf}(t) \quad (4)$$

The sideband-free signal can be obtained by directly subtracting the artifacts from the signal:

$$S_{w,0}(t) + \sum_m S_{m,0}(t) = [S_w(t) + \sum_m S_m(t)] - [S_{w,sa}^{df}(t) + S_{w,sa}^{uf}(t)] \quad (5)$$

This is the approach adopted in the current paper. An alternative approach to the sidebands-free signal is by first extracting the phases of the water and artifacts,

$$\vartheta(t) = \arg\,[S_{w,0}(t) + S_{w,sa}^{df}(t) + S_{w,sa}^{uf}(t)]$$

and then removing the effects of the phases:

$$S_{w,0}(t) + \sum_m S_{m,0}(t) = S_w(t)e^{-i\vartheta_G(t)}$$

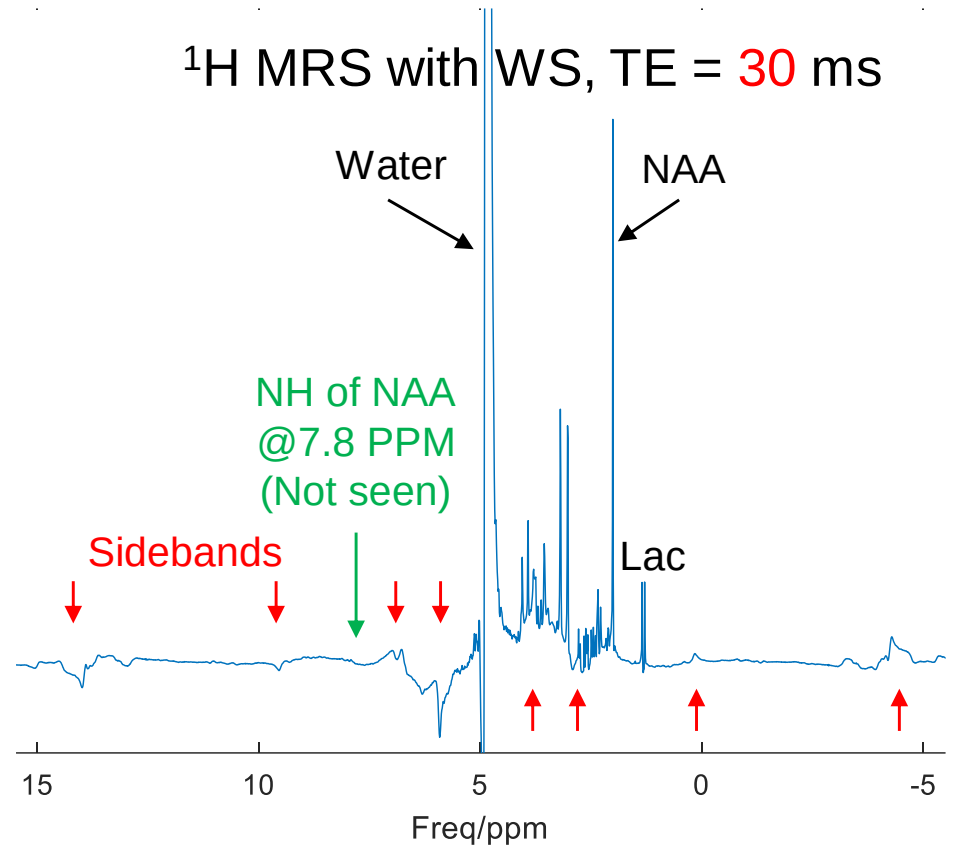

Figure 2. $^1$H MRS of the "Braino" phantom with water suppression. The resonances of NAA (7.82 ppm) and Cr (6.65 ppm) are not visible despite the very high SNR of the signal.

## 2.2 METHODS

### *2.2.1 Phantom MRSI Data acquisition*

We acquired single voxel phantom $^1$H MRS data on a 3T scanner (SIGNA$^{TM}$ Premier, GE Healthcare) using a 21-channel surface coil. The phantom is a spherical container ("Braino", GE Healthcare) that contains major brain metabolites with concentrations close to their normal physiological values (N-acetylaspartate: 12.5mM; creatine: 10 mM; choline: 3mM; glutamate: 12.5 mM; myo-inositol: 7.5mM; lactate: 5 mM). The data were acquired using a commercial PROBE-P sequence with the following parameters: TR/TE = 2000/30 ms, spectral width = 5000 Hz, FID points = 4096; number of excitations for the unsuppressed water = 16; number of excitations for water suppressed data = 16; voxel size = 4 x 4 x 4 cm$^3$. The center of the voxel was (4, 4, 4). WS was performed by the CHESS module embedded in the PROBE-P.

### 2.2.2 In vivo data acquisition

We acquired in vivo $^1$H MRSI data from a healthy volunteer using the same scanner and coil as used for the phantom experiments, following a protocol approved by the local institutional review board and after obtaining written informed consent from the volunteer. The MRSI data were acquired from a15 mm-thick slice whose lower side was 10 mm above the ACPC line. The MRSI data were acquired using the pulse sequence PROBE-P with the following parameters: FOV = 24 x 24 cm$^2$, number of phase encodings = 16 x 16 (nominal size of the voxel = 1.5 x 1.5 x 1.5 cm$^3$); TR/TE = 1500/68 ms; spectral width = 5000 Hz, datapoints in the FID = 4096. A water suppressed dataset and a non-water-suppressed dataset were acquired sequentially. Water suppression was achieved using the CHESS module imbedded in the sequence.

### 2.2.3 Data processing

All data processing procedures were implemented in Matlab R2019a (The MathWorks, Natick, MA).

*Pre-processing*. Major pre-processing steps include (1) weighted combination of multichannel coil data using the signal-to-noise ratio as the weighting factor;[24] (2) lineshape corrections in the time domain by dividing the unsuppressed water signal and multiplying a Lorentzian signal with a decay rate according to the FWHM of the water signal, which was extracted using the Matrix-Pencil based method (MPM);[25] and (3) water removal from the data.

*Sideband removal*. Sidebands in downfield were first recognized by the MPM-derived frequencies higher than the upper bound of water frequency, $f_{wu}$, which was set to 20 Hz (~ 0.15 ppm) for phantom data and 70 Hz (~ 0.55 ppm) for in vivo data. The sideband signal in the downfield was then reconstructed from the parameters of the sideband components according to the following equation,

$$S_{w,sa}^{df}(t) = \sum_{n=1}^{N} a_n e^{i2\pi f_n t - \alpha_n t + \varphi_n}$$

where $a_n$, $f_n$, $\alpha_n$, *and* $\varphi_n$ are the amplitude, frequency, decay rate, and phase of the *n*th component, respectively, and $N$ is the total number of the components.[14, 25] The sidebands in the upfield, $S_{w,sa}^{u}$, was obtained by taking the negative conjugate of $S_{w,sa}^{d}$, as expressed in Eq. 3. The sideband-free signal was obtained by subtracting the sidebands from the signal without WS, according to Eq. 5.

### 2.2.4 Performance evaluations

We used visual inspection and analytical comparisons to evaluate the performance of the method on both phantom and in vivo data. For visual inspection, we examined the real and the imaginary spectra before and after RIPS procedure, and we also compared the real and the imaginary spectra after RIPS with those acquired with WS. Special attention was paid to the overall baseline and the spectral segments with severe sideband artifacts. We used the SNR and linewidth of NAA as a metric of analytical comparison, where the noise level was calculated as the standard deviation of data from the signal free regions of the real spectra.

## 3 RESULTS

### 3.1 Phantom data

The RIPS method provides excellent results in removing strong sidebands from both the real part and the imaginary part of the phantom spectrum. As shown in Figure 3, the original NWS spectra were severely distorted by the sidebands from about -1500 Hz to about 1500 Hz on a 3T scanner. After the RIPS procedure, the sidebands were not visible except very tiny residues of the strongest sidebands at ±1000 Hz, which is well beyond the spectral region of interest. The baseline of the real part spectrum is flat, in contract with the WS spectrum, which suffers from the residue of sidebands resulted from incomplete suppression of the water signal (Figure 4).

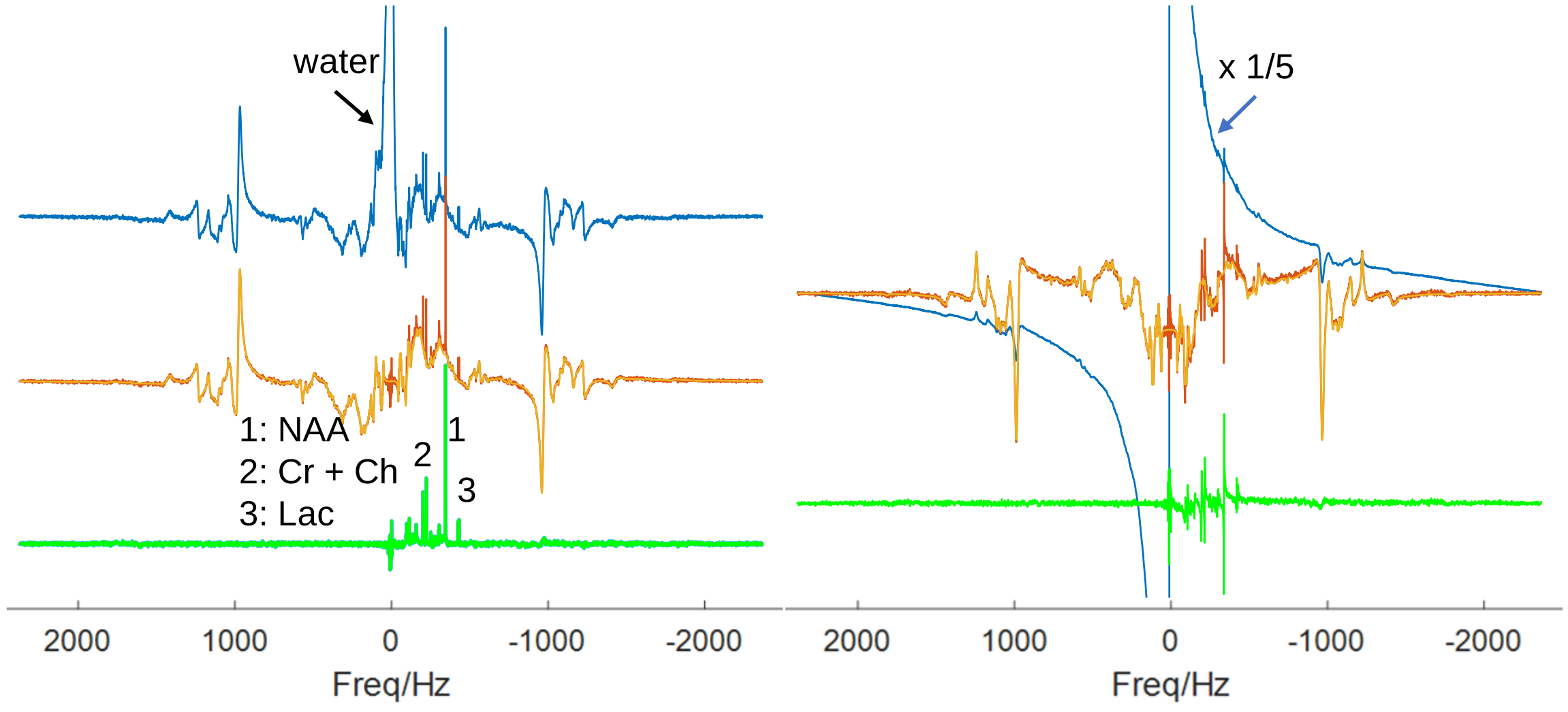

Figure 3. An example of phantom spectra (real part on the left-hand panel and imaginary on the right-hand panel) showing the results of major processing procedures. From top to bottom are the spectra of after eddy current correction (blue), the spectra after removing water (red) overlaid with the reconstructed sidebands (orange), and the clean spectra after removing sidebands. Note that the imaginary spectrum in blue is scaled by a factor of 1/5 and the frequency axis is in points.

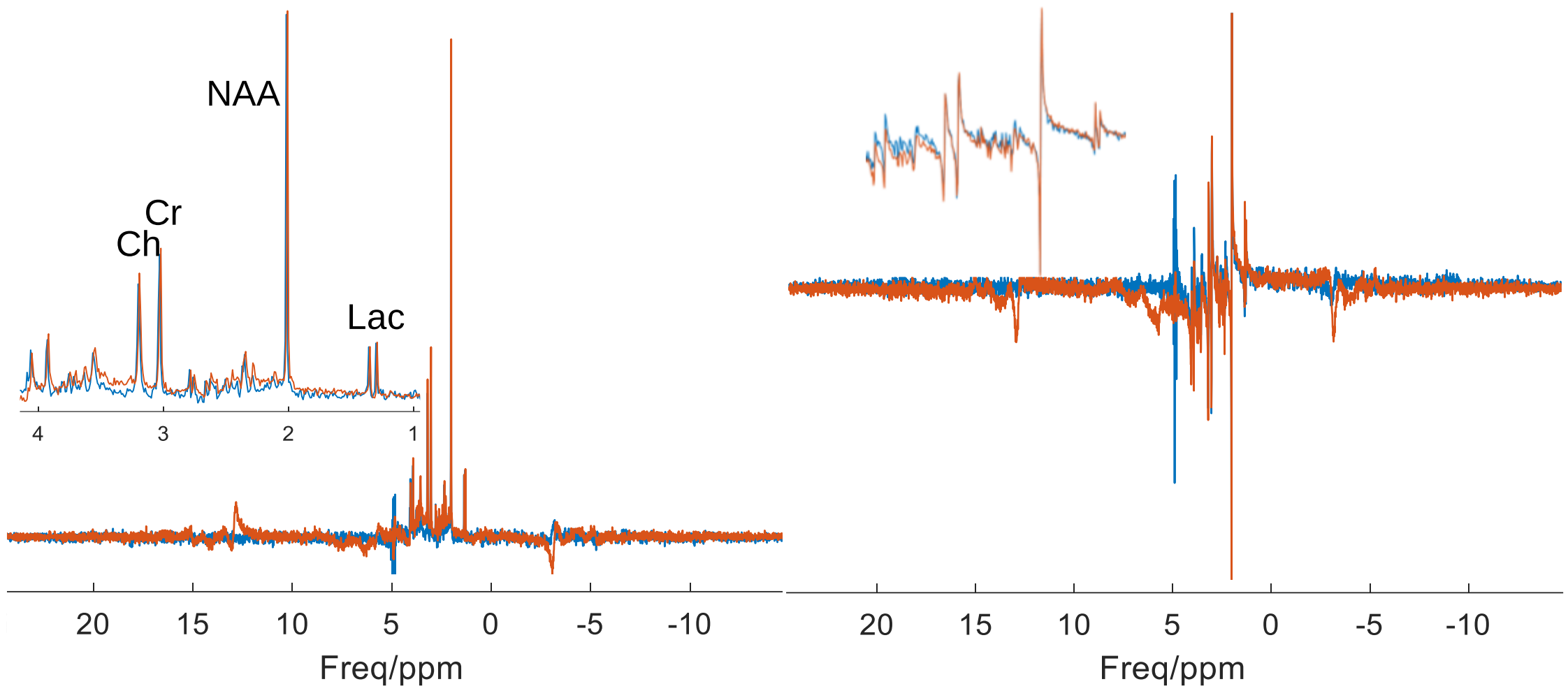

Figure 4. A comparison of NWS phantom spectra after removing sidebands (blue) and phantom WS spectra (red) in their real mode (left-hand panel) and imaginary mode (right-hand panel). The inset figures show the spectra in the range of 1 to 4 ppm. Note that the WS spectra exhibit residue sidebands due to insufficient water suppression.

## 3.2 In vivo data

The detailed results of the major data processing steps of the RIPS method and a comparison the RIPS result with a corresponding WS spectrum are shown in Figure 5. Compared with the phantom data, the in vivo data showed different sideband patterns and intensities, because the in vivo data were acquired at a longer TE, with a smaller voxel size, and at a different voxel location. The water signal was fitted very well as exhibited by the small water residue in the difference spectrum. After the RIPS procedure, both the upfield and downfield spectra show flat baseline and nearly completely removed sidebands, as seen in the water-free and signal-free regions. The pattern of the water residue (the green line in Figure 5) was changed from the difference (Diff) spectrum because of negative conjugate operation and subtraction of the spectra around the zero-frequency point. The spectrum after RIPS is highly like the spectrum acquired with WS. The Ch peak at 4.05 ppm disappeared and the Cr peak at 3.93 ppm were well retained in the NWS spectrum, while these peaks were partially suppressed in the WS spectrum. For numerical comparison, the original linewidths (LW) of water peaks in the WS and NWS spectra are 6.07 Hz and 6.10 Hz, respectively. The LW of the NWS water after lineshape correction is 5 Hz. The noise levels of the sidebands removed spectrum in the (6.0ppm, 24.2ppm) and (-2.5ppm, -14.9ppm) regions are 155 and 164, respectively, while the noise levels of the WS spectrum in the same two regions are 194 and 190, respectively. In the sideband removed spectrum, the LW and SNR of NAA are 10.01 Hz and 74.3, respectively, while they are 11.7 Hz and 54.1, respectively, in the WS spectrum, where the SNRs were calculated using noise levels of 164 and 190, respectively. More examples are shown in the matrix of voxels in the slice (Figure 6), in which the RIPS resultant spectra exhibit similar features.

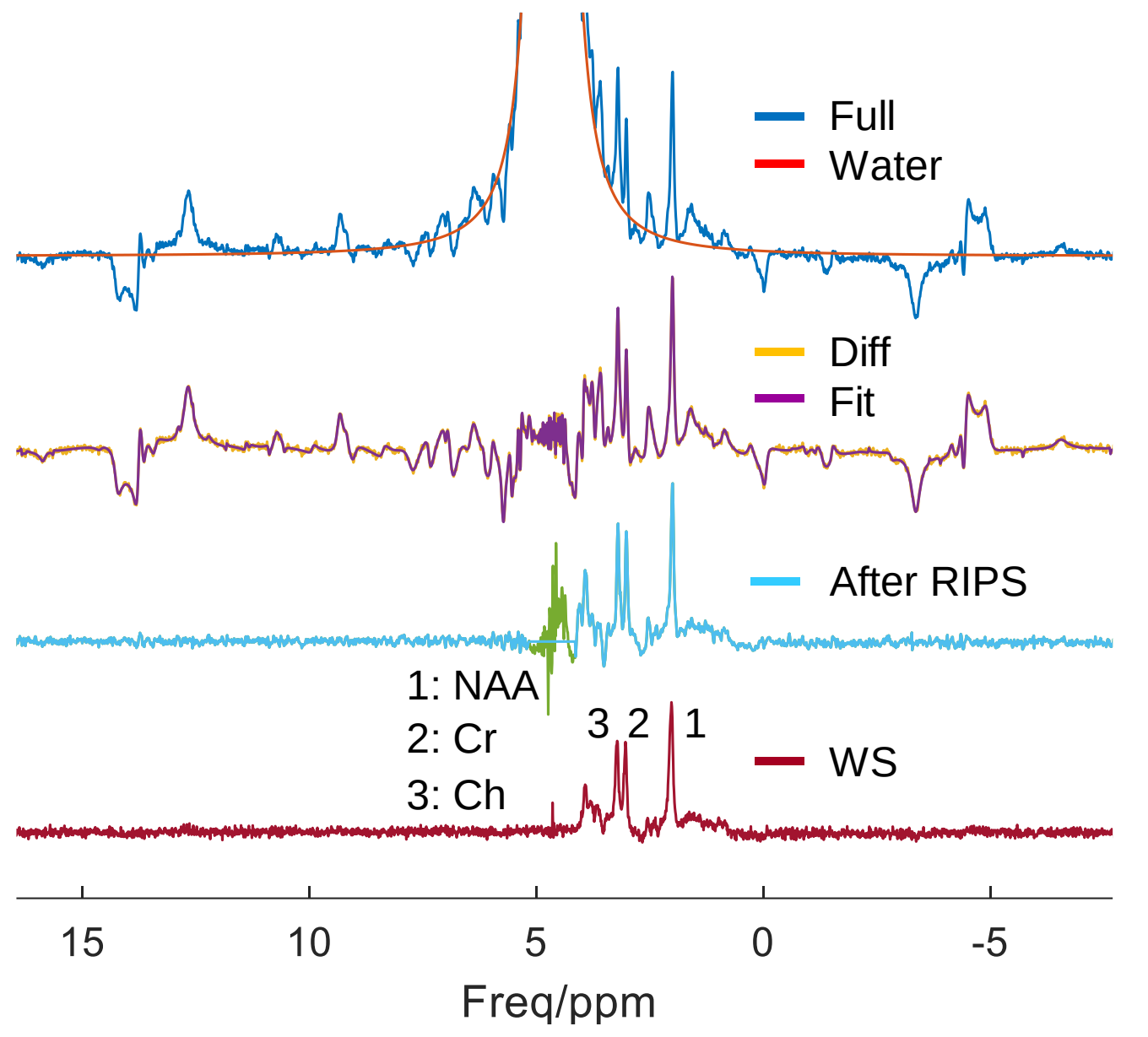

Figure 5. The processing a non-water-suppressed spectrum and comparison with water suppressed spectrum. From top to bottom: (1) Lineshape corrected spectra of the full signal and the water component, (2) The difference spectrum after removing water from the full signal and its fit, (3) The clean spectrum after eliminating the sidebands and the residue of water, and (4) the water suppressed spectrum from the same voxel.

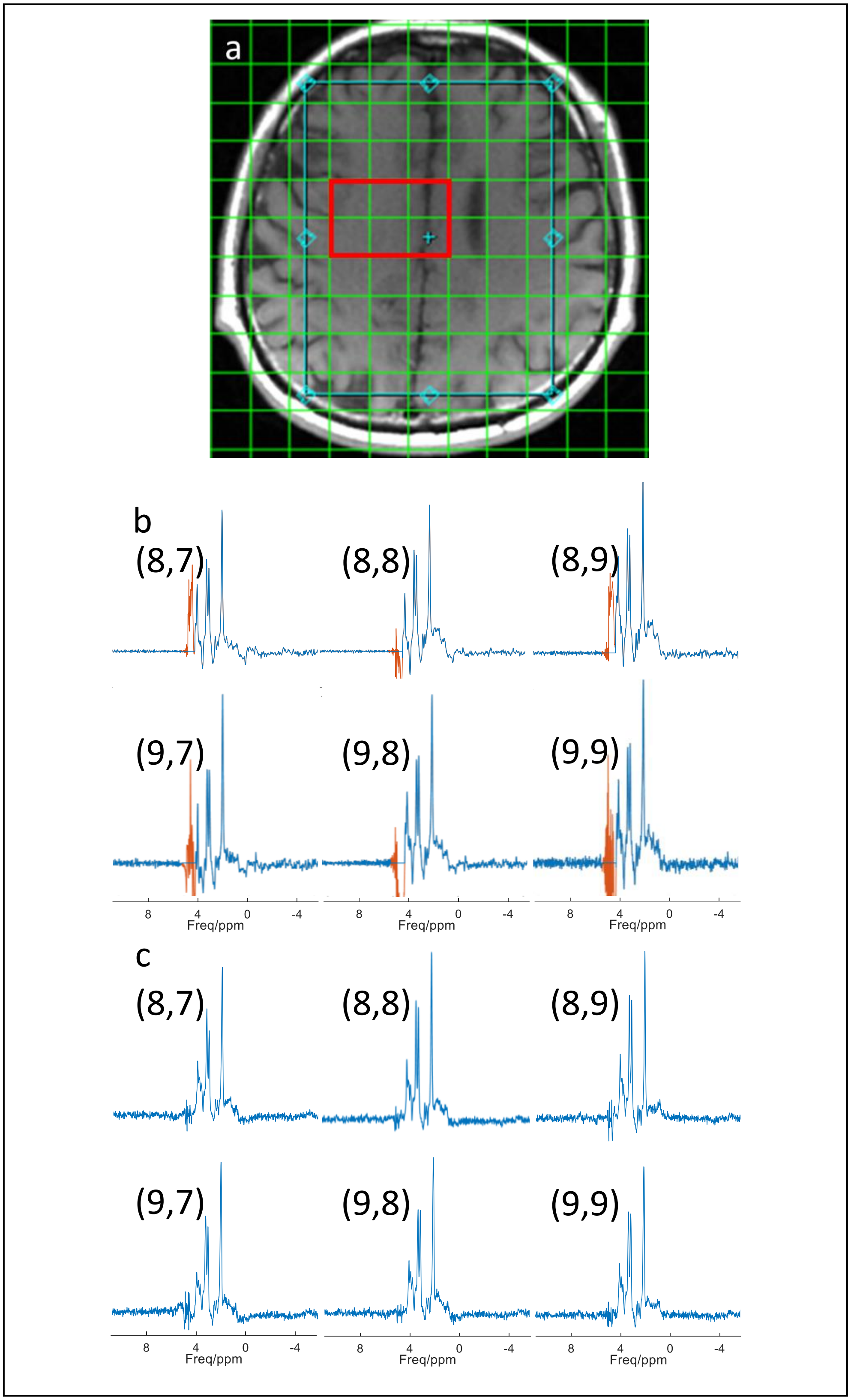

Figure 6. Comparison of the NWS spectra (b) and WS spectra (c) from a 2 x 3 voxel region shown in the red box with MRSI grid overlaid on the $T_1$-weighted structural image (a). The red lines in (b) are the water residues, which are removed for better visualization of the metabolite spectra.

## 4 DISCUSSIONS

We have presented a method for eliminating sideband artifacts in $^1$H MRS signal acquired without WS, and we demonstrated the performance of the method using both phantom and in vivo data. The method is based on different phase symmetries of the real and the imaginary parts of the sidebands.

A complete knowledge of the symmetric feature of sidebands is crucial for the application of phase symmetry-based sideband removal methods. Previously, the sidebands were described in numerous publications as being anti-symmetric around the water signal, and this was incorrectly considered to be true for both the real and the imaginary parts of the sidebands.[2, 11, 12, 22] As we have shown, the anti-symmetric feature holds only for the real part of the sidebands, while the imaginary part of the sidebands is symmetric. When the anti-symmetric property is incorrectly applied to sideband removal in the postprocessing, the sidebands in the real part are removed, but the sidebands in the imaginary part are doubled. This will lead to the following systematic or random errors in the measurement of metabolites, depending on whether the complex signal or only the real part of the signal is used in the spectral fitting. (1) If the complex signal is used in the spectral fitting, the nulled artifacts in the real part and doubled artifacts in the imaginary part will produce systematic errors comparable to those without removing the artifacts at all (see a Monte Carlo verification in Supplementary Materials). The sizes of the systematic errors depend on factors such as the amplitudes and phases of the sideband artifacts,[10] and there is no way to eliminate this systematic error. (2) If only the real part is used in the spectral fitting, that is, only ½ information is used, the random error of the measurement will increase by a factor of $\sqrt{2}$; to offset this increase in error, the scan time must be doubled.

The symmetric characteristics of the sidebands can be theoretically understood using an analytical expression of the sidebands.[2] The time domain MRS signal can be modeled as the sum of complex, exponentially decaying sinusoids:

$$s(t_n) = \sum_{m=1}^{M} A_m e^{-\alpha_m t_n + i\omega_m t_n + i\vartheta_m} + \varepsilon(t_n) \qquad (6)$$
$$t_n = (\mathrm{n} + \delta)\Delta\mathrm{t}, \mathrm{n} = 0, 1, 2, \ldots, (\mathrm{N} - 1)$$

where $M$ is the total number of signal components, $N$ is the number of datapoints, $\boldsymbol{A}$, $\alpha$, $\omega$, and $\vartheta$ are constants for amplitude, normalized decay, circular frequency, and phase, respectively, $t_n$ and $\Delta\mathrm{t}$ are the discrete sampling timepoint and sampling interval, respectively, and ε is the Gaussian noise. The real part of the MRS signal is symmetric and is termed as absorption spectrum in the frequency domain; the imaginary part of the signal is asymmetric and is termed as dispersion spectrum. The symmetry of the sideband artifacts, however, is exactly opposite to that of the MRS signal. An analytical expression of the sidebands is given as:[2]

$$S_{w,sa}(t) = S_{w,0}(t) \cdot 2i \sum_{l=1}^{L} J_1(\tilde{A}_{G,l}) \cos(\varphi_{G,l}) \quad (7)$$

where $J_1$ is the first-order Bessel function of the first kind, $\tilde{A}_{G,l}$ and $\varphi_{G,l}$ are variables related to the amplitudes and frequencies of the sidebands, respectively, and $L$ is the number of sideband artifacts. The terms in the summation symbol "Σ" in the above equation are real numbers. In the ideal case when water is on resonance, i.e., $\omega_w = 0$, the $S_{w,0}(t)$ is a real-number signal (its imaginary part is zero), and $S_{w,sa}(t)$ is an imaginary signal. We showed (Supplementary Materials) that the fact that the time domain sideband signal is imaginary determines the symmetry features of its spectra in frequency domain, which are anti-symmetric in the real part spectrum and symmetric in the imaginary part spectrum. We further note that the analytical feature of the sidebands in Eq. 7 agrees with that of the empirical expression of $S_{w,sa}(t)$ in Eq. 4, where its real part is zero by the observation-based definition in Eq. 3.

In addition to the complete knowledge about the symmetry of the sidebands, the absence of observable metabolite signals in the downfield region is a crucial prerequisite of the method (Figure 2). Some metabolites, such as the NH groups of NAA, Cr, and PCr, and the $NH_2$ group of Gln have resonances in the downfield at 7.82, 6.65, 7.30 ppm, and 6.82 ppm, respectively.[23] These resonances, if excited, will be observable and overlap with sideband artifacts in the downfield, making extraction of the sidebands impossible. In conventional $^1$H MRS experiments, however, these resonances are not excited using RF pulses with limited bandwidth and shifted center frequency.[26, 27] Therefore, the prerequisite of the proposed method is automatically satisfied.

The performance of the method depends on the accurate modeling and extraction of the water signal, which is 3-4 orders of magnitude larger than metabolite signals. The present paper used for water extraction an SVD-based matrix pencil method (MPM),[25, 28] which decomposes the signal into a series of exponentially decaying sinusoids and allows the identification of water components or baseline components by defining their frequencies and their decay rates, respectively.[29] This method worked very well for high quality data with narrow linewidths and little to mild lineshape distortion as manifested by small water residues and flat and symmetric baselines around water frequency (Figures 3-5). When the spectrum is severely broadened and distorted by field inhomogeneity and eddy current as in the short TE, the modeling of water signal by this method may not be accurate due to incomplete account of water components within the water range or the interference of artificial spectral components from outside of the water range.[29] As a result, the spectrum after water removal will leave asymmetric baselines around the center of the spectrum. This is detrimental to the RIPS, since the unbalanced baseline will not be canceled but be doubled in the upfield after the RIPS procedures. We adopted the following remedial measures to mitigate the problem: (1) we increased the range of water to ±0.7 ppm, (2) we used a large number of components (~150) in MPM to over fit the entire spectrum and to reconstruct water signal from the fitting signal components within the water range, (3) we performed lineshape correction by deconvolving the spectrum with the water signal, and (4) we extracted water components again from the lineshape corrected spectrum, using an iterative procedure so that all MPM determined signal components fell within the range of water. These measures improved the performance of the RIPS. A side effect of the lineshape correction is the variation of the noise levels in the RIPS spectra, which were resulted from different linewidths in the lineshape correction that matched their original linewidths (Fig. 6).

Residual sidebands and noise level are two metrics of the methods for $^1$H MRSI without WS. We classify the errors in the RIPS processed data into the system error caused by the residual sidebands and the random error caused by noise that consists of the original noise in the upfield and the noise from reflected downfield signal. We consider the following three cases. (1) In the worst case, the entire downfield signal, which include sidebands and the noise, is reflected by RIPS to the upfield; the sidebands and, therefore, the system errors are removed, and the random noise level is increased to $\sqrt{2}\sigma$, where $\sigma$ is the noise level in the original signal that should be the same in both upfield and downfield. (3) In the ideal case, only the sidebands $S_{w,sa}^{df}$in the downfield is accurately determined and is reflected by the RIPS to the upfield; the sidebands are removed, and the noise level remains to be $\sigma$. (3) In the real-world case, the fitting of the sidebands in the downfield is impaired by the noise, and the errors have two sources. One is $dS_{w,sa}^{df}$, which is the error in the sidebands fitting: $S_{w,sa}^{df} \pm dS_{w,sa}^{df}$. For this error, the following relation holds $\sigma_{CRLB} < dS_{w,sa}^{df} \ll \sigma$, where $\sigma_{CRLB}$is the Cramer-Rao Lower Bound that is usually much smaller than the noise level. The other error is the false sidebands resulting from the noise. This error can be in the level of $\sigma$, but the number of the false sidebands is small. This error can be further reduced by applying prior knowledge of the sidebands to modeling of the sidebands, e.g., to the number of singular value components when an-SVD-based method is used. In the numerical example given in the in vivo results (Figure 5), the ratio of the noise levels in the upfield to the downfield is 164/155, or 1.058. the 5.8% increase of the upfield noise can be attributed to the errors caused by the RIPS. Overall, the system errors and random errors caused by the RIPS are minimal. In contrast, reference-based methods may have larger, visible system errors due to the mismatch of the sidebands in the reference and the in vivo signals,[18-20] while the reference-free methods may have larger random noise.[11]

The lineshape correction has multiple functions in improving the quality in the RIPS process. Different from simple spectral linebroadening, which improves the SNR at the expense of spectral resolution, the deconvolution-based lineshape correction can improve the SNR or the spectral resolution, or both. The numerical results for Figure 5 show that, both the SNR and spectral resolution of NAA were simultaneously improved for the RIPS spectrum as compared with the WS spectrum, with 74.3 vs 54.1 10.0 Hz vs 11.7 Hz for SNR and resolution, respectively. On the other hand, the SNR may be traded off for the spectral resolution, as revealed by increased noise level in the last voxels in Figure 6. Lineshape correction also corrected eddy current caused lineshape distortion and corrected frequency shift. The latter is vitally important for the frequency aligning of the reflected sidebands with the original sidebands in the upfield.

The present work has some limitations. First, it only focused on the description of the method and the demonstration of its effectiveness in removing sideband artifacts in $^1$H MRSI data acquired without WS. The effectiveness of the method is shown by visually comparing the resultant spectra with the WS spectra and by numerically comparing the SNR of the resultant spectra with those of the spectra with WS, where the noise levels were calculated in the metabolite signal-free regions in both the real and the imaginary parts. It did not directly compare the performance of the present method with those of the previously published methods. However, the results in a previous comparison show that the current method is not inferior to those methods.[21] Second, the method was applied to $^{1}$H MRSI acquired at a TE of 68ms. Applying this method to short TE data is desirable and plausible. But the short TE $^{1}$H MRSI data suffers from stronger sideband artifacts and severer lineshape distortion. This requires methods to accurately model the water signal without baseline distortion and phase shift in the residual spectrum.

## 5 CONCLUSION

The RIPS method allows the removal of sideband artifacts from both the real and the imaginary parts of the signal. This method can be applied to $^{1}$H MRSI data acquired with any pulse sequences without WS and does not require any reference scans. The improvement of the proposed method in removal sidebands artifacts and reducing noise depends on complete extraction of the water signal and accurate modeling of the sidebands in the downfield.

1. Alger JR. Quantitative proton magnetic resonance spectroscopy and spectroscopic imaging of the brain: a didactic review. *Top Magn Reson Imaging* 2010; **21**(2)**:** 115-128.

2. Dong Z. Proton MRS and MRSI of the brain without water suppression. *Prog Nucl Magn Reson Spectrosc* 2015; **86-87:** 65-79.

3. Kreis R, Ernst T, Ross BD. Development of the human brain: in vivo quantification of metabolite and water content with proton magnetic resonance spectroscopy. *Magn Reson Med* 1993; **30**(4)**:** 424-437.

4. Ishihara Y, Calderon A, Watanabe H, Okamoto K, Suzuki Y, Kuroda K *et al.* A precise and fast temperature mapping using water proton chemical shift. *Magn Reson Med* 1995; **34**(6)**:** 814-823.

5. Kuroda K. Non-invasive MR thermography using the water proton chemical shift. *Int J Hyperthermia* 2005; **21**(6)**:** 547-560.

6. Dong Z, Kantrowitz JT, Mann JJ. Improving the reproducibility of proton magnetic resonance spectroscopy brain thermometry: Theoretical and empirical approaches. *NMR Biomed* 2022**:** e4749.

7. Dong Z, Milak MS, Mann JJ. Proton magnetic resonance spectroscopy thermometry: Impact of separately acquired full water or partially suppressed water data on quantification and measurement error. *NMR Biomed* 2022; **35**(6)**:** e4681.

8. Posporelis S, Coughlin JM, Marsman A, Pradhan S, Tanaka T, Wang H *et al.* Decoupling of Brain Temperature and Glutamate in Recent Onset of Schizophrenia: A 7T Proton Magnetic Resonance Spectroscopy Study. *Biol Psychiatry Cogn Neurosci Neuroimaging* 2018; **3**(3)**:** 248-254.

9. Zhang Y, Taub E, Mueller C, Younger J, Uswatte G, DeRamus TP *et al.* Reproducibility of whole-brain temperature mapping and metabolite quantification using proton magnetic resonance spectroscopy. *NMR Biomed* 2020; **33**(7)**:** e4313.

10. Clayton DB, Elliott MA, Leigh JS, Lenkinski RE. 1H spectroscopy without solvent suppression: characterization of signal modulations at short echo times. *J Magn Reson* 2001; **153**(2)**:** 203-209.

11. Serrai H, Clayton DB, Senhadji L, Zuo C, Lenkinski RE. Localized proton spectroscopy without water suppression: removal of gradient induced frequency modulations by modulus signal selection. *J Magn Reson* 2002; **154**(1)**:** 53-59.

12. Dong Z, Dreher W, Leibfritz D. Experimental method to eliminate frequency modulation sidebands in localized in vivo 1H MR spectra acquired without water suppression. *Magn Reson Med* 2004; **51**(3)**:** 602-606.

13. Dreher W, Leibfritz D. New method for the simultaneous detection of metabolites and water in localized in vivo 1H nuclear magnetic resonance spectroscopy. *Magn Reson Med* 2005; **54**(1)**:** 190-195.

14. Dong Z, Dreher W, Leibfritz D. Toward quantitative short-echo-time in vivo proton MR spectroscopy without water suppression. *Magn Reson Med* 2006; **55**(6)**:** 1441-1446.

15. Emir UE, Burns B, Chiew M, Jezzard P, Thomas MA. Non-water-suppressed short-echo-time magnetic resonance spectroscopic imaging using a concentric ring k-space trajectory. *NMR Biomed* 2017; **30**(7).

16. Pruessmann KP, Weiger M, Scheidegger MB, Boesiger P. SENSE: sensitivity encoding for fast MRI. *Magn Reson Med* 1999; **42**(5)**:** 952-962.

17. Lustig M, Donoho D, Pauly JM. Sparse MRI: The application of compressed sensing for rapid MR imaging. *Magn Reson Med* 2007; **58**(6)**:** 1182-1195.

18. Chadzynski GL, Klose U. Proton CSI without solvent suppression with strongly reduced field gradient related sideband artifacts. *MAGMA* 2013; **26**(2)**:** 183-192.

19. Özdemir MS, Deene YD, Fieremans E, Lemahieu I. Quantitative proton magnetic resonance spectroscopy without water suppression. *Journal of Instrumentation* 2009; **4**(06)**:** P06014-P06014.

20. Chadzynski GL, Klose U. Chemical shift imaging without water suppression at 3 T. *Magn Reson Imaging* 2010; **28**(5)**:** 669-675.

21. Dong Z, Liu F, Li M, Milak M, Jambawalikar S. A comparison of reference-based methods for removing artifacts in non-water-suppressed $^1$H MRSI data. *The Joint Annual Meeting ISMRM-ESMRMB*: Paris, France. , 2018, p 1332.

22. Elliott MA, Clayton DB, Lenkinski RE. $^1$H spectroscopy without water suppression: Removal of sideband modulations at short TE. *The Joint Annual Meeting ISMRM-ESMRMB*: Glasgow, Scotland, United Kingdom, 2001, p 1667.

23. Govindaraju V, Young K, Maudsley AA. Proton NMR chemical shifts and coupling constants for brain metabolites. *NMR Biomed* 2000; **13**(3)**:** 129-153.

24. Dong Z, Peterson B. The rapid and automatic combination of proton MRSI data using multi-channel coils without water suppression. *Magn Reson Imaging* 2007; **25**(8)**:** 1148-1154.

25. Lin Y-Y, Hodgkinson P, Ernst M, Pine A. A Novel Detection-Estimation Scheme for Noisy NMR Signals: Applications to Delayed Acquisition Data. *J Magn Reson* 1997; **128**(1)**:** 30-41.

26. Fichtner ND, Giapitzakis IA, Avdievich N, Mekle R, Zaldivar D, Henning A *et al.* In vivo characterization of the downfield part of (1) H MR spectra of human brain at 9.4 T: Magnetization exchange with water and relation to conventionally determined metabolite content. *Magn Reson Med* 2018; **79**(6)**:** 2863-2873.

27. Povazan M, Schar M, Gillen J, Barker PB. Magnetic resonance spectroscopic imaging of downfield proton resonances in the human brain at 3 T. *Magn Reson Med* 2022; **87**(4)**:** 1661-1672.

28. Hua Y, Sarkar TK. Matrix pencil method for estimating parameters of exponentially damped/undamped sinusoids in noise. *IEEE Trans Acoust, Speech, Sig Process* 1990; **38**(5)**:** 814-824.

29. Dong Z, Dreher W, Leibfritz D, Peterson BS. Challenges of using MR spectroscopy to detect neural progenitor cells in vivo. *AJNR Am J Neuroradiol* 2009; **30**(6)**:** 1096-1101.

# Supplementary Materials

Real and imaginary part symmetries (RIPS) and artifacts removal in $^1H$ magnetic resonance spectroscopy without water suppression

Zhengchao Dong[1,2]

1. Department of Psychiatry, Columbia University, College of Physicians and Surgeons, New York, NY, USA

2. New York State Psychiatric Institute, New York, NY, USA.

# 1 Spectral symmetries of real or imaginary signal

## 1.1 A complex time domain signal and its spectrum in the frequency domain

A simulated, complex NMR signal with a frequency of -200 Hz and a small zero-order phase is shown in Figure S1. The complex time domain signal consists of sinusoids with an exponential decay, and the complex frequency domain signal consists of absorption and dispersion peaks with a phase error.

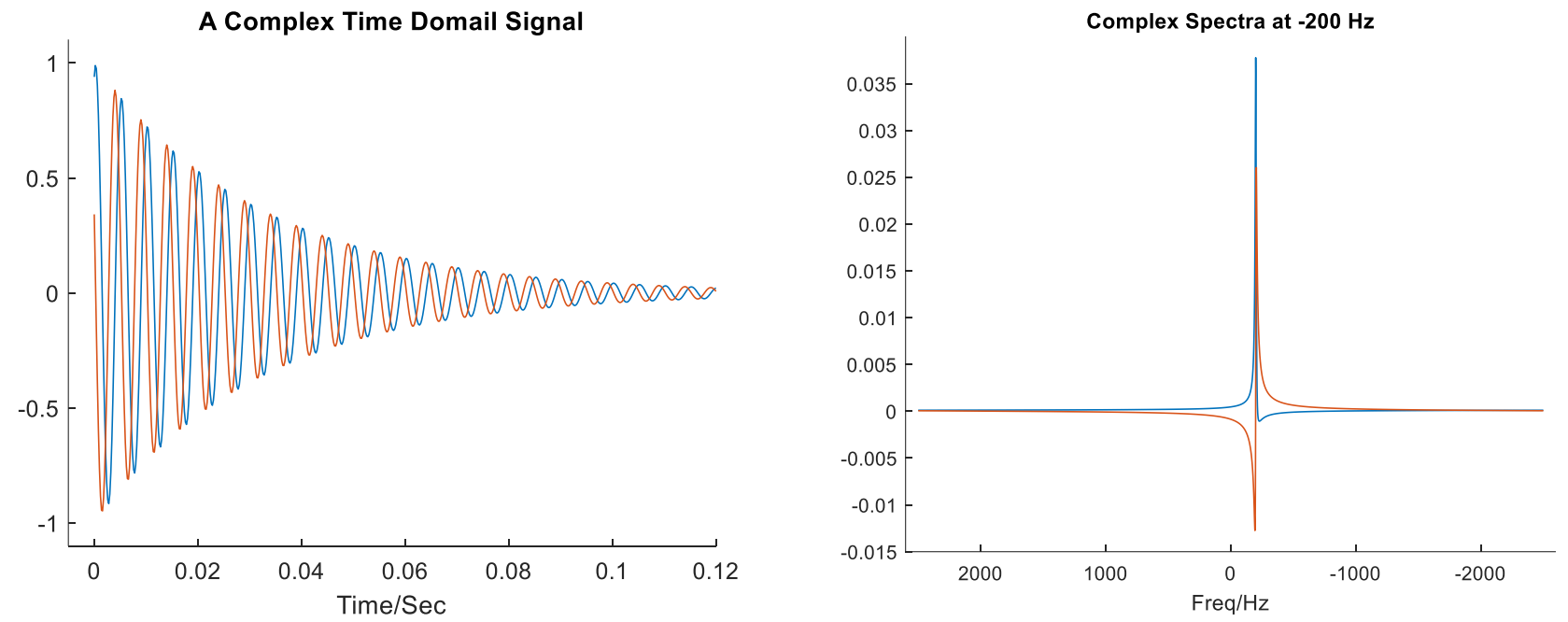

**Figure S1**. A complex time domain signal (left-hand panel) and its spectrum in the frequency domain (right-hand panel). When the signal is complex, its Fourier transform produces a spectrum with correct frequencies at -200 Hz. Note: the complex time domain signal has a small zero-order phase so that it has a non-zero imaginary part (the red line), and the real and imaginary parts of the frequency domain spectra have small zero-phase distortion (the blue and red lines, respectively).

## 1.2 Spectra of a real- or an imaginary-time domain signal

When the above time domain signal has only a real part or an imaginary part, its frequency domain signal will be a complex spectrum with peaks at ±200 Hz, and the symmetries of the two peaks depends both on the parts of the spectrum and on the parts of the time domain signal.

(1) If the time domain signal is real, the two peaks of its real part spectrum are symmetric, and the two peaks of its imaginary part spectrum are anti-symmetric. And
(2) If the time domain signal is imaginary, the two peaks of its real part spectrum are anti-symmetric, and the two peaks of its imaginary part spectrum are symmetric.

As shown in the literature[1], the time domain sideband artifacts are imaginary, the frequency domain sidebands will be anti-symmetric for their real part and symmetric for their imaginary part.

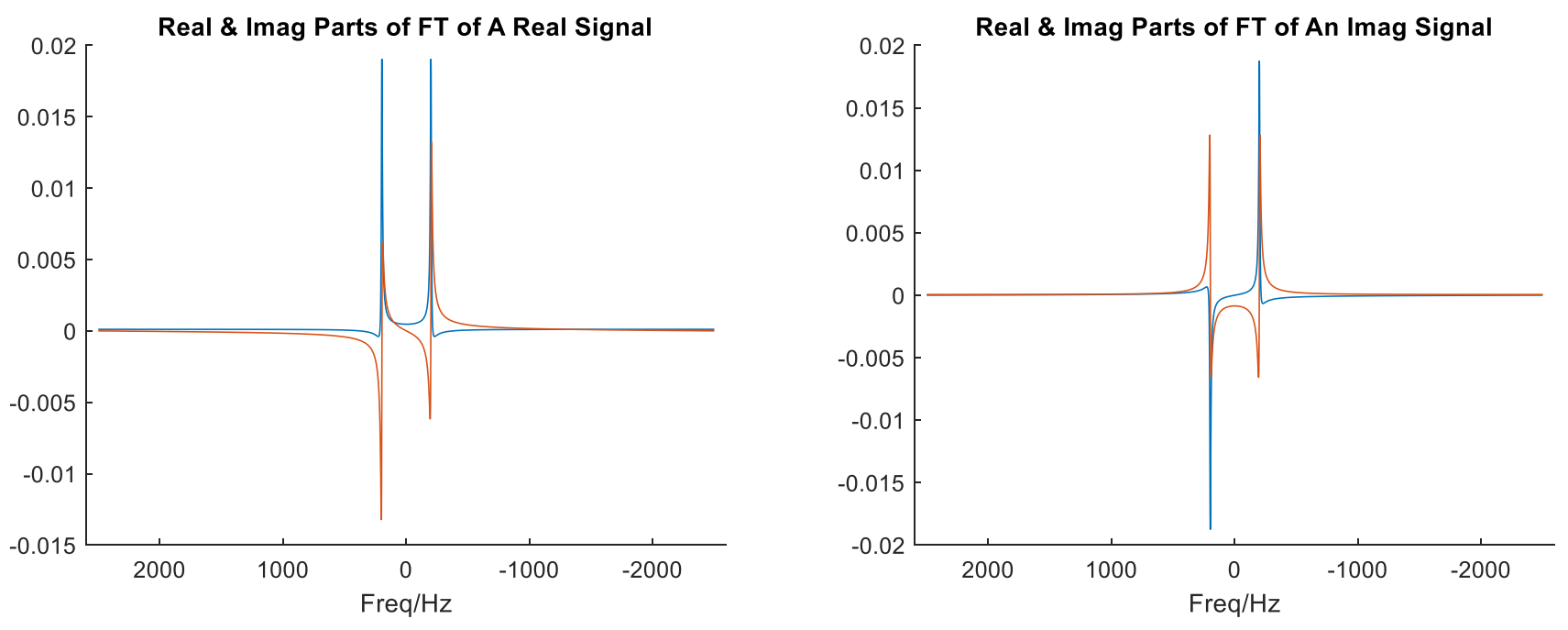

**Figure S2**. Real and imaginary parts of the Fourier transform of a real signal (left-hand panel) and an imaginary signal (right-hand panel). Note: (1) For non-complex signal, its spectrum has two peaks at ± 200 Hz; (2) for the FT of a real signal, the real part of the spectrum is symmetric, and the imaginary part is anti-symmetric; and (3) for the FT of an imaginary signal, the real part of the spectrum is anti-symmetric, and the imaginary part is symmetric.

# 2 Monte Carlo verification of the spectral fitting errors of the NWS spectra corrected using only the anti-symmetric property of the sidebands

## 2.1 Test signals

The model of the test signals is given as follows,

$$S_n(t) = aS_0(t)e^{-\pi bt+i2\pi ft+i\varphi} + \varepsilon_n(t) \qquad \text{Eq.S1}$$
$$n = 1, 2, \dots, N$$

where $S_0$ is the signal of a single metabolite, $a$, $b$, $f$, and $\varphi$ are values for amplitude scaling, linebroadening, frequency shift, and phase shift of the signal, respectively, $\varepsilon$ is the added Gaussian noise, and $n$ is the $n$th of the total $N$ Monte Carle tests.

In this study, we used the simulated NAA signal, normalized its time domain amplitude to 1, for $S_0$. The values for amplitude scaling, linebroadening, frequency shift, and phase shift are $a$ = 10 (a.u.), $b$ = 7 Hz, $f$ = 1 Hz, and $\varphi = 0$. The noise level is 0.25, which makes the frequency domain SNR of NAA, defined as the CH3 peak height to the standard deviation of the noise, be about 97. The number of tests $N$ = 1000. An example of the test signals is shown in Figure S3, with the sideband artifacts given in Eq. S2.

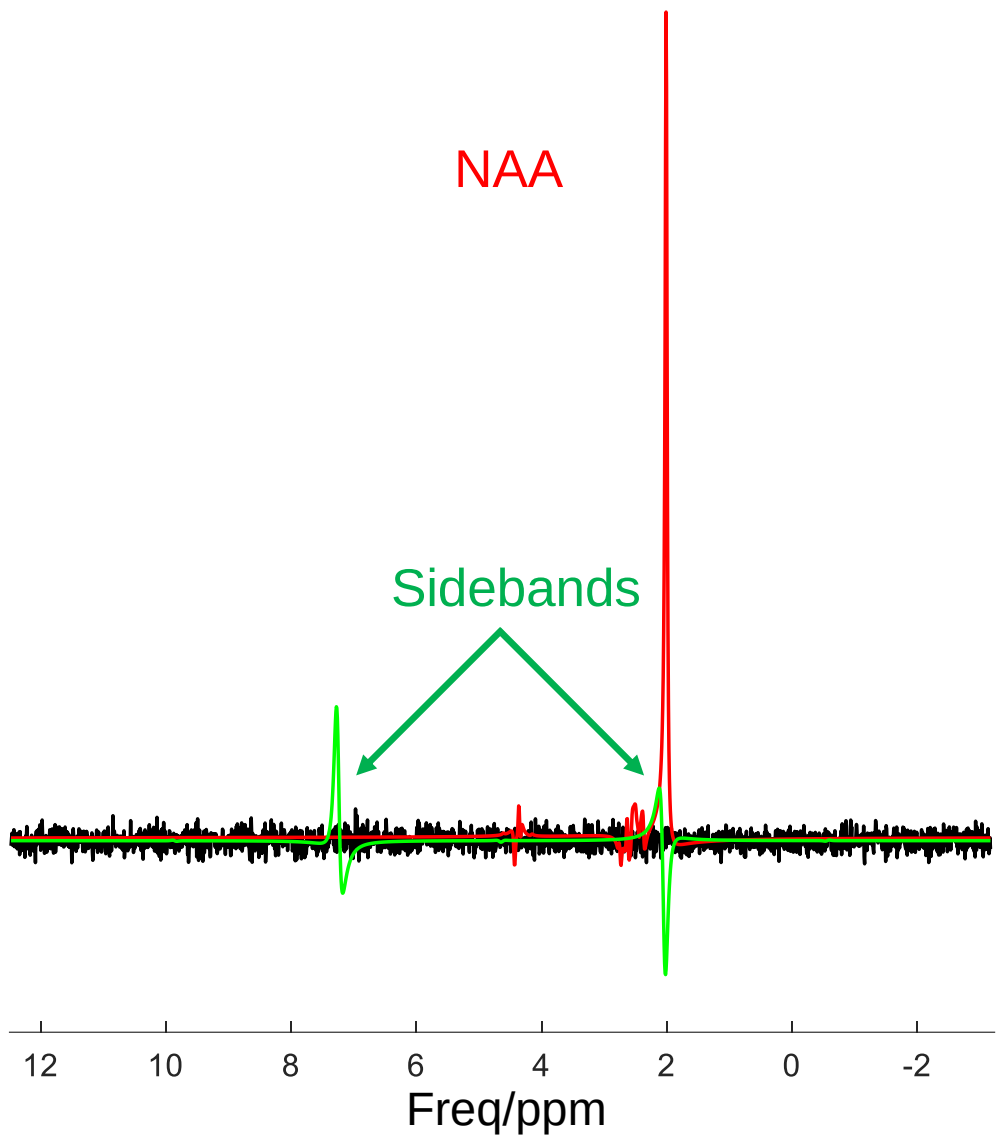

**Figure S3.** An example of the test signals. Shown in the figure are the NAA signal (red), one of the three artifact signals (green), and one of 1000 noise signals (black). Note that the phases of the sideband artifacts vary with the phase $\varphi_G$ in Eq. S2.

## 2.2 Sideband artifacts

The phase factor of the sideband artifacts in Eq. 1 is given as

$$\vartheta_G = 2\pi t \cdot A_G cos(2\pi f_G t + \varphi_G) \quad \text{Eq. S2}$$

where $A_G = 2$, $f_G = 333$ Hz, $\varphi_G = 0, \pi/2, \pi/2$, respectively, for 3 sets of sideband artifacts. We generated the sideband artifacts, $S_{ar}$, as follow,

$$S_{ar} = S_{on} e^{i\vartheta_G} \quad \text{Eq. S3}$$

where $S_{on}$ is an on-resonance carrier signal with an amplitude of 2000 (a.u.) and a linewidth of 10 Hz. In real-world experiment, the carrier signal is the on-resonance water signal.

## 2.3 Monte Carlo study

We performed Monte Carlo (MC) studies for the following four signals:

(i) The signal $S_n$ in Eq. S1, which has no sideband artifacts and where $S_0$ is simulated NAA signal,

(ii) The signal in Eq. S1 added with 2-fold of the imaginary part of the sideband artifacts $S_{ar}$ in Eq. S3,

(iii) The sum of $S_n$ and artifact signal $S_{ar}$, and

(iv) The real part of the signal $S_n$ without artifacts.

To facilitate the MC studies, we generated N = 1000 sets of complex, Gaussian noise signals with the same standard deviation and with the same number of data points as that of the above (i) – (iv) signals. Each of the (i) – (iv) signals was added with the 1000 nois signals to form a set of 1000 test signals for an MC study.

We used the NAA signal $S_0$ as the basis signal to fit the N = 1000 test signals for the four parameters in Eq. S1, which are amplitude scaling $a$, linebroadening $b$, frequency shift $f$, and phase shift $\varphi$. We calculated the means, the differences with the corresponding given values, and the standard deviations of the fitting parameters.

## 2.4 Results

The results of the Monte Carlo studies for the amplitudes are given in Table S1. For signals without sideband artifacts, the relative systematic error, defined as the relative difference between the mean estimated amplitudes and the given value, is in the order of $10^{-4}$. The random error of the estimate, defined as the standard deviation (SD) of the estimated amplitudes, is 4.76 x $10^{-2}$, which is close to the Cramer-Rao lower bound (CRLB) of 5.24 x $10^{-2}$.

For the signal added with 2-fold imaginary sidebands, the relative systematic errors are sizeable but vary with the phases of the sidebands; but the random errors of the estimates agree well with the CRLB.

For signals spoiled with the sidebands, the systematic errors are also sizeable for and vary with the phases of the sidebands; the random errors agree well with the CRLB.

Finally, when only the real part of the signal is used in the fitting, the system error is still small, but the random error is about √2 times of those of the other three signals or the CRLB.

The results not only verified the theoretical analyses in the main text but also revealed that the systematic errors caused by the sideband artifacts, or the unremoved (doubled) imaginary artifacts may differ for signals stemming from different voxels in SI because of different phases of the artifacts.

**Table S1**. The means, differences with the given value of 10 a.u., and standard deviations of the estimated amplitude scales: Mean/Diff/SD, for the four different signals each with three different artifacts (if added).

| Signal | Amplitude Scale (a = 10) | | |
|---|---|---|---|
| | 1. ($\varphi_G = 0$) | 2. ($\varphi_G = \pi/4$) | 3. ($\varphi_G = \pi/2$) |
| (*i*) Without artifacts | 9.998/-1.46x$10^{-3}$/0.0476 | = | = |
| (*ii*) With 2X imag. artifacts | 14.371/4.37/0.0531 | 12.689/2.689/0.0595 | 9.0578/-0.942/0.0575 |
| (*iii*) With artifacts, fit complex signal | 9.437/-0.562/0.0501 | 8.130/-1.870/0.0453 | 8.0479/-1.952/0.0434 |
| (*iv*) Without artifacts, fit real part signal | 9.984/-0.0158/0.0723 | = | = |

Note: "=" means that the values equal those of case 1.

1. Dong Z. Proton MRS and MRSI of the brain without water suppression. *Prog Nucl Magn Reson Spectrosc* 2015; **86-87:** 65-79.